\documentclass[%
 reprint,
 superscriptaddress,
 amsmath,
 amssymb,
prl,
]{revtex4-1}

\usepackage{fullpage}
\usepackage[colorlinks   = true, urlcolor     = cyan, linkcolor    = blue, citecolor   = blue]{hyperref}
\usepackage{bm}
\usepackage{url}

\newcommand{\avg}[1]{{\left\langle{#1}\right\rangle}}
\newcommand{\dd}{\text{\,}\mathrm{d}}

\begin{document}

\title{A note on transversal flexoelectricity in two-dimensional systems}

\author{David Codony}
\affiliation{College of Engineering, Georgia Institute of Technology, Atlanta, GA 30332, USA}
\affiliation{Laboratori de C\`{a}lcul Num\`{e}ric, Universitat Polit\`{e}cnica de Catalunya, Barcelona, E-08034, Spain}

\author{Irene Arias}
\affiliation{Centre Internacional de Mètodes Numèrics en Enginyeria (CIMNE), 08034 Barcelona, Spain}
\affiliation{Laboratori de C\`{a}lcul Num\`{e}ric, Universitat Polit\`{e}cnica de Catalunya, Barcelona, E-08034, Spain}

\author{Phanish Suryanarayana}
\email{phanish.suryanarayana@ce.gatech.edu}
\affiliation{College of Engineering, Georgia Institute of Technology, Atlanta, GA 30332, USA}

\date{\today}

\begin{abstract}
In a recent letter \cite{Springolo2021PRL}, the authors introduced the \emph{effective flexoelectric} coefficient $\mu^\textrm{2D}$ for quantifying the flexoelectric effect in 2D systems, and reported a disagreement with the \emph{flexoelectric} coefficient $\mu_\text{T}$ introduced in Ref.~\cite{Codony2021PRM}, attributed to the neglect of the \emph{metric term} $\varphi^\text{M}$ --- quadrupolar moment of the unperturbed charge density --- in the formulation of Ref.~\cite{Codony2021PRM}. Here, we show that the model in Ref.~\cite{Codony2021PRM} is correct and is in agreement with that in Ref.~\cite{Springolo2021PRL}. The discrepancies in the numerical values of the coefficients arise due to the difference in their definitions: $\mu_\textrm{T}$ measures changes in bending-induced out-of-plane polarization, whereas $\mu^\textrm{2D}$ measures changes in bending-induced voltage drop across the 2D system.
\end{abstract}

\maketitle

Considering a slab in $\mathbb{R}^3$ under uniform bending deformation with curvature $\kappa$, the electric displacement  can be written in the undeformed black(flat) configuration as \cite{Stengel2013, Codony2021JMPS, Springolo2021PRL}:
\begin{align}
D_Z =J\epsilon_0E_Z+P_Z,\label{DZ}
\end{align}
where $J{}=1+\kappa Z$ is the Jacobian of the transformation, $\epsilon_0$ is the vacuum permittivity, $E_{Z}$ is the electric field, and $P_Z$ is the  polarization, with the subscript $Z$ used to denote the component perpendicular to the flat 2D system. Thereafter, the macroscopic electric displacement is obtained as:
\begin{align}
\avg{D_Z} =\epsilon_0\avg{E_Z}+\kappa\epsilon_0\varphi^\text{M}+\avg{P_Z}, \label{avdDZ}
\end{align}
where $\avg{.}$ denotes the surface area averaged volume integral, and the metric term: 
\begin{align}
\varphi^\text{M} = \avg{Z E_Z} = -\frac{1}{2} \avg{Z^2 \frac{\partial E_Z}{\partial Z}} \approx -\frac{1}{2 \epsilon_0} \avg{Z^2 \rho_0}, \hspace{-2mm}
\end{align}
with $\rho_0 = J \rho$ being the nominal charge density, i.e.~the electronic+ionic charge per unit undeformed volume. The second equality is obtained using integration by parts, with the resulting term  then approximated to leading order by substituting Eq.~\eqref{DZ} in Gauss's law for electrostatics written in the reference configuration \cite{Stengel2013, Codony2021JMPS, Springolo2021PRL}: $\partial(J \epsilon_0 E_Z)/\partial Z = \rho_0$. 

The constitutive model for the macroscopic polarization in slabs under uniform bending takes the form \cite{Codony2021JMPS, Codony2021PRM}:
\begin{equation}\label{const}
\avg{P_Z}=\mu_\text{T} J^2 \kappa  + J \epsilon_0(\epsilon_{ZZ}-1)\avg{E_Z} \,,
\end{equation}
where $\epsilon_{ZZ}$ is the $ZZ$ component of the relative permittivity. On averaging over the thickness and considering its linear response $\delta_\kappa := \partial/\partial \kappa | _{\kappa=0}$, Eq.~\eqref{const} becomes: 
\begin{align}\label{dkconst}
\delta_\kappa\!\avg{P_Z} &=
\mu_\text{T} + \epsilon_0(\epsilon_{ZZ}-1)\delta_\kappa\!\avg{E_Z}
,
\end{align}
which is consistent with the definition of $\mu_\text{T}$ in Ref.~\cite{Codony2021PRM}:
\begin{align}\label{mut}
\mu_\text{T}=\delta_\kappa\!\avg{P_Z}|_{\avg{E_Z}=0}.
\end{align}
Considering Eq.~\eqref{dkconst} and the linear response of Eq.~\eqref{avdDZ}, we obtain:
\begin{align}
\label{avdDconst}
\delta_\kappa\!\avg{D_Z} = \mu_\text{T} + \epsilon_0\varphi^\text{M} + \epsilon_0\epsilon_{ZZ}\delta_\kappa\!\avg{E_Z} \,,
\end{align}
from which it follows that:
\begin{align}\label{rel}
\mu^\textrm{2D}=\epsilon_0\delta_\kappa V=-\epsilon_0\delta_\kappa\!\avg{E_Z}|_{\avg{D_Z}=0}=\frac{\mu_\text{T}+\epsilon_0\varphi^\text{M}}{\epsilon_{ZZ}},
\end{align}
with $V$ being the difference in electrostatic potential across the 2D system.

Given that for a freestanding 2D system embedded in infinite vacuum, $\epsilon_{ZZ}\rightarrow1$, the equation above can be written as:
\begin{align}\label{St}
\mu^\textrm{2D}=\mu_\text{T} +\epsilon_0\varphi^\text{M},
\end{align}
which matches Eq.~(4) in Ref.~\cite{Springolo2021PRL}. This shows that the flexoelectric coefficient $\mu_\textrm{T}$ in Ref.~\cite{Codony2021PRM} and the effective flexoelectric coefficient $\mu^\textrm{2D}$ in Ref.~\cite{Springolo2021PRL} derive from the same mathematical framework and are both correctly defined, the former referring to the rate of bending-induced out-of-plane polarization, the latter to the rate of bending-induced voltage drop, both with respect to curvature. The metric term relates both quantities, as it is needed to compute voltage drop, but not polarization. Eq.~\eqref{St} is the zero curvature limit of the voltage drop rate in bending deformation settings at finite $\kappa$, see Appendix for details.

Overall, the above analysis suggests that both $\mu_\textrm{T}$ and $\mu^\textrm{2D}$ are correctly defined, their relationship is well established, and are physically sound but different measures of flexoelectricity in 2D systems.

\section{Appendix: Voltage drop across 2D systems under finite uniform bending}

Adopting the bending deformation map of a flat slab in the $X$--$Y$ plane around the $Y$--axis \cite{Stengel2013, Codony2021JMPS, Springolo2021PRL}, Poisson's equation in real space $(x,y,z)$ for electrostatics
\begin{align}
    \nabla^2 \phi(\mathbf{x}) = -\frac{\rho(\mathbf{x})}{\epsilon_0}
\end{align}
can be written in the undeformed (flat) configuration coordinates $(X,Y,Z)$ \cite{Stengel2013, Codony2021JMPS, Springolo2021PRL} as
\begin{align}\label{pois}
    \nabla_0\cdot\big[J\mathbf{C}^{-1}\cdot\nabla_0 \phi(\mathbf{X}) \big] = -J\frac{\rho(\mathbf{X})}{\epsilon_0},
\end{align}
where $\phi$ is the electrostatic potential, $\rho$ is the total (electronic+ionic) charge density, $\nabla_0=\mathbf{F}^\textrm{T}\cdot\nabla$ is the gradient operator in the flat configuration, in terms of the deformation gradient $\mathbf{F}$; $J=\textrm{det}(\mathbf{F})=1+\kappa Z=\kappa r$ is the Jacobian of the deformation, with $r=Z+\kappa^{-1}$ denoting the radial coordinate in the bent configuration; and
\begin{align}
\mathbf{C}^\textrm{-1}=\mathbf{F}^\textrm{-1}\cdot\mathbf{F}^\textrm{-T}=\begin{bmatrix}
J^{-2} & 0 & 0 \\
0 & 1 & 0 \\
0 & 0 & 1
\end{bmatrix}
\end{align}
is the inverse of the metric tensor. For a uniform bending deformation, Eq.~\eqref{pois} simplifies thus to
\begin{align}\label{si}
\frac{\partial}{\partial X}\left[\frac{1}{J}\frac{\partial \phi}{\partial X} \right]
+
\frac{\partial}{\partial Y}\left[J\frac{\partial \phi}{\partial Y} \right]
+
\frac{\partial}{\partial Z}\left[J\frac{\partial \phi}{\partial Z} \right]
=-J\frac{\rho}{\epsilon_0}.
\end{align}
The integral of the equation above along $X$ and $Y$ directions is
\begin{align}\label{in}
\int\!\!\!\!\int_{X\!Y}\frac{\partial}{\partial Z}\left[J(Z)\frac{\partial \phi(\mathbf{X})}{\partial Z} \right]\!\!\dd S
=\!\int\!\!\!\!\int_{X\!Y} -J(Z)\frac{\rho(\mathbf{X})}{\epsilon_0}\dd S,
\end{align}
given that the integrals of the first two terms in Eq.~\eqref{si} vanish due to $\nabla\phi(X,Y,Z)$ being periodic along $X$ and $Y${. The integral of the first term in Eq.~\eqref{si} vanishes since
\begin{align}\nonumber
&\int\!\!\!\!\int_{X\!Y}\frac{\partial}{\partial X}\left[\frac{1}{J(Z)}\frac{\partial \phi(\mathbf{X})}{\partial X} \right]\!\!\dd S
\\\nonumber
&={}
\frac{1}{J(Z)}\int_{Y}\left[
\int_X \frac{\partial^2 \phi(\mathbf{X})}{\partial X^2} \dd X
\right]\!\!\dd Y
\\\nonumber
&={}
\frac{1}{J(Z)}\int_{Y}\left[
\frac{\partial \phi(X^+,Y,Z)}{\partial X}
-
\frac{\partial \phi(X^-,Y,Z)}{\partial X}
\right]\!\!\dd Y
\\
&= 0.
\end{align}
The integral of the second term in Eq.~\eqref{si} analogously vanishes.

Using $\dd Z = \dd r$ and $J=\kappa r$, Eq.~\eqref{in} can be written in radial coordinates as
\begin{align}\label{rad}
\frac{1}{r}\frac{\partial}{\partial r}\left[ r\frac{\partial}{\partial r}\overline{\phi}(r) \right]
=-\frac{\overline{\rho}(r)}{\epsilon_0},
\end{align}
where
\begin{subequations}\begin{align}
    \overline{\phi}(r) & = \frac{1}{S}\!\int\!\!\!\!\int_{X\!Y}\phi(X,Y,r-\kappa^{-1}) \dd S,\\
    \overline{\rho}(r) & = \frac{1}{S}\!\int\!\!\!\!\int_{X\!Y}\rho(X,Y,r-\kappa^{-1}) \dd S
\end{align}\end{subequations}
are the surface-averaged electrostatic potential and charge density, with $S$ being the surface area.
The radial Laplacian operator in Eq.~\eqref{rad} has an associated Green's function of the form
\begin{align}
G(r,r')=\frac{1}{2}\left|\,\log\!\left(\frac{r'}{r}\right)\right|,
\end{align}
and the averaged electrostatic potential is written as
\begin{align}
\overline{\phi}(r)
&=\nonumber
\int_{-\infty}^{+\infty}-G(r,r')\frac{\overline{\rho}(r')}{\epsilon_0}r'\dd r'
\\{}&=
-\frac{1}{2\epsilon_0}
\int_{R_1}^{R_2}\left|\,\log\!\left(\frac{r'}{r}\right)\right|\overline{\rho}(r')r'\dd r',
\end{align}
where we used that $\overline{\rho}(r)$ is localized in $R_1<r<R_2$.

The voltage drop across the 2D system is
\begin{align}\nonumber
V &= \overline{\phi}(R_2) - \overline{\phi}(R_1)
=
\frac{1}{\epsilon_0}
\int_{R_1}^{R_2}\log\!\left(r\right)\overline{\rho}(r)r\dd r
\\{}&
=
\frac{1}{\epsilon_0}
\int_{Z_1}^{Z_2}\frac{\log\!\left(J\right)}{J-1}\,\overline{\rho}_0(Z)Z\dd Z,
\end{align}
where charge-neutrality of the system is used, and the last equality is expressed in the flat configuration coordinates with $\overline{\rho}_0=J\overline{\rho}$ being the surface-averaged electronic+ionic charge per unit undeformed volume.

The bending-induced voltage drop rate at a given $\kappa$ is
\begin{align}\label{op}\nonumber
\hspace{-0.86em}\frac{\partial V}{\partial \kappa}
&=
-\frac{1}{\epsilon_0}
\int_{Z_1}^{Z_2}\left(\frac{\log(J)}{J-1}-\frac{1}{J}\right)\frac{\overline{\rho}_0(Z)}{\kappa} Z\dd Z
\\{}&{}\hphantom{{}=}
+
\frac{1}{\epsilon_0}
\int_{Z_1}^{Z_2}\frac{\log\!\left(J\right)}{J-1}\,\frac{\partial\,\overline{\rho}_0(Z)}{\partial \kappa}Z\dd Z,
\nonumber\\\nonumber
&=
-\frac{1}{\epsilon_0}
\!\int_{Z_1}^{Z_2}\!\!\frac{1}{\kappa^2}\!
\left[
\log(1+\kappa Z) - \frac{\kappa Z}{1 + \kappa Z}
\right]
\overline{\rho}_0(Z)\dd Z
\\{}&{}\hphantom{{}=}
+
\frac{1}{\epsilon_0}
\int_{Z_1}^{Z_2}\!\frac{1}{\kappa}\bigg[\log(1+\kappa Z)\bigg]\,\frac{\partial\,\overline{\rho}_0(Z)}{\partial \kappa}\dd Z,
\end{align}
and the limiting case for the flat 2D system is
\begin{align}\nonumber
\lim_{\kappa\rightarrow 0}\frac{\partial V}{\partial \kappa}
={}&\nonumber-\frac{1}{2\epsilon_0}
\int_{Z_1}^{Z_2}\overline{\rho}_0(Z) Z^2\dd Z
\\{}&{}\nonumber
+
\frac{1}{\epsilon_0}
\int_{Z_1}^{Z_2}\frac{\partial \,\overline{\rho}_0(Z)}{\partial \kappa}Z\dd Z
\\={}&{}\nonumber
\varphi^\textrm{M}+\frac{\delta_\kappa\!\avg{P_Z}}{\epsilon_0}
\\={}&{}\frac{\mu^\textrm{2D}}{\epsilon_0},
\end{align}
where the terms of Eq.~\eqref{op} in square brackets have been expanded in a Taylor series around $\kappa=0$.

\vspace{11em}

The metric term is thus needed to compute the bending-induced voltage drop rate at $\kappa =0$, termed \emph{effective flexoelectric} coefficient $\mu^\textrm{2D}$ in \cite{Springolo2021PRL}, but not the flexoelectric coefficient $\mu_\text{T}$, as defined in \cite{Codony2021PRM}.

\emph{Acknowledgments.} 
The authors acknowledge helpful discussions with M.~Stengel.
D.C. acknowledges the support of the Spanish Ministry of Universities through the Margarita Salas fellowship (European Union-NextGenerationEU).
D.C.~and I.A.~acknowledge the support of the Grant CEX2018-000797-S funded by MCIN/AEI/10.13039/501100011033.
I.A. acknowledges the support of the European Research Council (StG-679451) and Generalitat de Catalunya (2017-SGR-1278 and the ICREA Academia award). P.S.~acknowledges the support of the U.S. National Science Foundation (CAREER-1553212).

\bibliographystyle{apsrev4-1}
%

\end{document}